\documentstyle[12pt,epsfig]{article}

\newcommand{\resetcounter}{\setcounter{equation}{0}}     
\newcommand{\non} {\nonumber}
\newcommand{\ti} {\tilde}

\begin{document}

\thispagestyle{empty}
\begin{titlepage}
\begin{flushright}
HUB-EP-96/3 \\
hep-th/9601130 \\
Januar 1996
\end{flushright}
\vspace{0.3cm}
\begin{center}
\Large \bf T-S-T Dual Black Hole 
\end{center}
\vspace{0.5cm}
\begin{center}
Bj\"orn Andreas $^{\hbox{\footnotesize{1}}}$ ,  \\
{\sl Institut f\"ur Physik, Humboldt--Universit\"at,\\
 Invalidenstrasse 110, D--10115 Berlin, Germany}
\end{center}
\vspace{0.6cm}

\begin{abstract}
\noindent
The sequence of intertwined T-S-T duality transformations acting on the
4D static uncharged black hole leads to a new black 
hole background with horizon and singularity exchanged. It is
shown that this space-time is extendible too. In particular
we will see that a string moving into a black hole is dual to a
string leaving a white hole. That offers the possibility that
a test-string does not see the singularity.
\end{abstract}

\vspace{0.3cm}
\footnotetext[1]{E-MAIL:andreas@qft2.physik.hu-berlin.de}
\vfill
\end{titlepage}


\setcounter{page}{1}

\resetcounter

The sequence of intertwined T-S-T duality transformations was first
proposed by I.Bakas [\ref{bakas}]. He argued that if one performs T-S-T
duality transformations on a pure gravitational background with one
Killing symmetry, one obtains a new background which will always be
pure gravitational as well. Moreover, it turns out that T-S-T can be
considered as a SL(2,${\bf R}$) transformation in the space of 
string background metrics. This SL(2,${\bf R}$) coincides with the
action of the Ehlers-Geroch SL(2,${\bf R}$) symmetry group of vacuum
Einstein spaces upon reduction from four to three dimensions [\ref{ehlers}]
provided that there is at least one Killing symmetry.\\
It is known that an exact conformal field theory describing a black 
hole in 2D space-time is an SL(2,${\bf R}$)/U(1) gauged WZW model 
[\ref{witt}].
T-duality acting on such a background is well understood and leads to
stringy properties of the black hole [\ref{giveon},\ref{luest}]. 
In the case of
a Euclidean space-time the so called semi-infinite cigar is transformed 
to an infinite funnel. For a Lorentzian space-time duality interchanges
different regions in the Kruskal-Szekeres coordinates[\ref{witt},
\ref{giveon}], especially
duality exchanges the horizon with the singularity and
the asymptotically flat space-time, which corresponds
to the Lorentzian cigar, with the space-time inside the black hole 
singularity. In particular it has been shown that the 2D black hole 
is self dual.\\
Dual geometries of 4D black holes were studied in
[\ref{Q1},\ref{Q2},\ref{Q3},\ref{Q4}]. In [\ref{Q1}] 
the T-duality transformation was applied to time translations of the 
Schwarzschild metric. The dual metric defined a geometry with naked
singularities at $r=0$ and $r=2M$ and remained a spherically symmetric
solution of the string background equations, but was not a black hole.
Further it was shown [\ref{Q1}] that a dual geometry with respect to the 
SO(3) symmetry, i.e. duality with respect to non-abelian isometries is
neither spherically symmetric nor asymtotically flat. \\
In the sequel we will show that the sequence of intertwined T-S-T duality
transformations acting on the 4D black hole solution produces a new 
spherically symmetric background that is asymtotically flat and has 
a singularity at $r=2M$ and a horizon at $r=0$.
In particular we will show that T-S-T duality transformation acts on
the 4D static black hole just like the T-duality on the 2D black hole.
Before starting with T-S-T let us recall that the action of T-duality 
for a nontrivial field configuration involving the metric, dilaton 
and antisymmetric tensor which are independent of the time coordinate 
is given by (for a review of T-duality [\ref{giveon2}]):

\begin{eqnarray}
\ti{G}_{00}&=&\frac{1}{G_{00}},\ \ \ \ti{G}_{0a}=\frac{B_{0a}}{G_{00}},\ \ \
 \ti{B}_{0a}=\frac{G_{0a}}{G_{00}}\non\\
\ti{G}_{ab}&=&G_{ab}-\frac{G_{0a}G_{0b}-B_{0a}B_{0b}}{G_{00}},\ \ \
\ti{B}_{ab}=B_{ab}-\frac{G_{0a}B_{0b}-G_{0a}B_{0b}}{G_{00}}\non\\ 
 \ti{\phi}&=&\phi-\frac{1}{2}logG_{00}\non  .
\end{eqnarray}

where '0' denotes the time direction.
In order for the dual geometries to give string vacua they have to satisfy
the string background equations to lowest order in $\alpha^{\prime}$ 
[\ref{giveon2},\ref{perry}].\\ \\ 
The 4D uncharged static black hole solution can be regarded 
as special case of a gravitational string background with zero 
dilaton $\phi$ and anti-symmetric tensor field $B_{\mu\nu}$. The isometry
group of the Schwarzschild metric is given by time translations together
with the SO(3) space rotations.\\ \\ The metric can be given in the form :

\begin{eqnarray*}
ds^{2}= -\left(1-\frac{2M}{r}\right)dt^{2}
        +\left(1-\frac{2M}{r}\right)^{-1}dr^{2}
        +r^{2}\left(d\theta^{2}+sin^{2}\theta d\phi^{2}\right).
\end{eqnarray*}

The metric is singular for $r=0$ and $r=2M$. One can show 
[\ref{hawking}] that $r=0$ is a real singularity but $r=2M$
just a coordinate singularity, reflecting deficiency in the used
coordinate system and therefore being removable. If one calculates the
Riemann tensor scalar invariant one finds: 

\begin{eqnarray*}
R_{\mu\nu\lambda\sigma}R^{\mu\nu\lambda\sigma}=\frac{10M^{2}}{r^{6}}.
\end{eqnarray*}

The scalar is finite at $r=2M$ and diverges for $r\rightarrow 0$.\\

{\bf\underline{T-duality:}}\\

If we perform a T-duality transformation with respect to the t-coordinate
we get the dual metric:

\begin{eqnarray*}
ds_{D}^{2}=-\frac{dt^{2}}{1-\frac{2M}{r}}+\frac{dr^{2}}{1-\frac{2M}{r}}
            +r^{2}d\Omega^{2}
\end{eqnarray*}

and $d\Omega^{2}=d\theta^{2}+sin^{2}\theta d\phi^{2}$. The dilaton is 
given by :

\begin{eqnarray*}
\phi_{D}=-\frac{1}{2} log\left(1-\frac{2M}{r}\right).
\end{eqnarray*}

Metric and dilaton solve the string-frame background equations.
If we switch to the Einstein-frame then the metric takes the form:
 
\begin{eqnarray*}
\bar{ds}_{D}^{2}=-dt^{2}+dr^{2}+(r^{2}-2Mr)d\Omega^{2}
\end{eqnarray*}

This metric defines a geometry
with naked singularities at $r=0$ and $r=2M$, as it was already 
pointed out [\ref{Q1}]. We can verify this by computing the dual 
scalar curvature

\begin{eqnarray*}
{\cal R}=\frac{2M^{2}}{(2M-r)^{2}r^{2}}  .
\end{eqnarray*}

It is instructiv to compute again the Riemann tensor scalar invariant
which is given by

\begin{eqnarray*}
R_{\mu\nu\lambda\sigma}R^{\mu\nu\lambda\sigma}=\frac{3M^{4}}
                         {(r^{2}-2Mr)^{4}}.
\end{eqnarray*}

A T-duality transformation of the 4D black 
hole gives us a spherically symmetric solution of the string background 
equations with two naked singularties which is not a black hole.\\

{\bf\underline{S-duality:}}\\

Now we perform a S-duality transformation [\ref{lu}].
We have $B_{\mu\nu}=0\rightarrow b=0$ and S-duality reduces to 

\begin{eqnarray*}
\phi\longrightarrow -\phi,
\end{eqnarray*}


and we set $-\phi = \hat{\phi}$. The metric in the Einstein-frame 
remains fixed under S-duality and is given by 

\begin{eqnarray*}
\hat{ds}_{D}^{2}=e^{-2\hat{\phi}_{D}}ds_{D}^{2} .
\end{eqnarray*}

If we go back to the string-frame, we will find the metric

\begin{eqnarray*}
ds^{2}_{D}=\left(1-\frac{2M}{r}\right)dt^{2}+(1-\frac{2M}{r})dr^{2}
+(r-2M)^{2}d\Omega^{2}
\end{eqnarray*}

which together with the dilaton

\begin{eqnarray*}
\hat{\phi}=\frac{1}{2}log(1-\frac{2M}{r})
\end{eqnarray*}

solves the equation of motion in the string-frame.\\

{\bf\underline{T-duality:}}\\

We perform finally a T-duality transformation with respect to the
t-coordinate. The T-S-T dual metric to the former Schwarzschild 
metric is then given by:

\begin{eqnarray*}
ds_{T-S-T}^{2}=-\left(1-\frac{2M}{r}\right)^{-1}dt^{2}+(1-\frac{2M}{r})dr^{2}
+(r-2M)^2 d\Omega^{2} .
\end{eqnarray*}

The dual dilaton vanishes:

\begin{eqnarray*}
\phi_{T-S-T}&=&\hat{\phi}-\frac{1}{2}log(1-\frac{2M}{r})\\
            &=&0 \ \ .
\end{eqnarray*}

It is easy to check that the equation of motion are satisfied 
by the T-S-T dual metric. The metric becomes singular at 
$r=0$ and $r=2M$ but a calculation of the Riemann scalar invariant
suggests that $r=0$ is not a real physical singularity, but rather
one which is a result of a bad choice of coordinates like $r=2M$
in the Schwarzschild solution. The invariant is given by:

\begin{eqnarray*}
R_{\mu\nu\lambda\sigma}R^{\mu\nu\lambda\sigma}=
\frac{10M^{2}}{(r-2M)^{6}} \ \ .
\end{eqnarray*}

The curvature invariant takes at $r=M$ the same value as the invariant
coming from the Schwarzschild metric.
Our new space-time is defined for $r<0$. Similar to the case
of the Schwarzschild solution we have to check if $r=0$ is 
a null hypersurface dividing the manifold into two disconnected
components:

\begin{eqnarray*}
\mbox{I}&:& -\infty < r < 0 \ \ ,\\
\mbox{II}&:& 0 < r < 2M \ \ .
\end{eqnarray*}

Inside region II the coordinates t and r reverse their
character ( t-spacelike, r-timelike).\\
To get a better understanding of our new solution we have to prove 
that our solution can be extended when r tends to 0.
I will follow the maximal extension procedure for the known
Schwarzschild solution [\ref{hawking}].
We start with a congruence of ingoing radial null geodesics
given by 

\begin{eqnarray*}
t=r-2Mln|r|+c \ \ ,
\end{eqnarray*}

and $t\rightarrow -t$ defines outgoing radial null geodesics. In 
the following we suppress the integration constant $c=2Mln|2M|$.
Now we change to a new time coordinate in which the ingoing 
geodesics become straight lines

\begin{eqnarray*}
t^{*}=t+2Mln|r| \ \ .
\end{eqnarray*}

If we differentiate this equation with respect to r and substituting
it for dt in the line element $ds_{T-S-T}^{2}$, we find a new 
line element that we call dual Eddington-Finkelstein line element

\begin{eqnarray*}
ds^{2}_{T-S-T}=-\left(1-\frac{2M}{r}\right)^{-1}
          (dt^{*})^2+\frac{4M}{r-2M}dt^{*}dr
         +\frac{r-4M}{r-2M}dr^{2}+(r-2M)^{2}d\Omega^{2} \ \ .
\end{eqnarray*}

This solution is regular for the whole range $0<r<2M$. Our transformation
$t\rightarrow t^{*}$ extends the coordinate range from $-\infty<r<0$ to
$-\infty<r<2M$. The time reversed solution for outgoing radial null
geodesics can be obtained by introducing another time coordinate
$t_{*}=t-2Mln|r|$. If we introduce an advanced and a retarded null 
coordinate

\begin{eqnarray*}
v=t^{*}-r \ \ , \ \ \ \ w=t_{*}+r \ \ ,
\end{eqnarray*}

the corresponding dual Eddington-Finkelstein metric becomes:

\begin{eqnarray*}
\mbox{advanced:}\ \ {ds}_{T-S-T}^{2}&=&
                  -\left(1-\frac{2M}{r}\right)^{-1}dv^{2}-2dvdr
                  +(r-2M)^{2}d\Omega^{2}\\
\mbox{retarded:}\ \ {ds}_{T-S-T}^{2}&=&
                  -\left(1-\frac{2M}{r}\right)^{-1}dw^{2}+2dwdr
                  +(r-2M)^{2}d\Omega^{2} \ \ .
\end{eqnarray*}

If we compare with the original advanced and retarded solutions 
coming from the Schwarzschild line element [\ref{hawking}] we will find 
the advanced (retarded) dual solution coincide at $r=M$ with the retarded
(advanced) original solution, latter are given by:

\begin{eqnarray*}
\mbox{advanced:}\ \ {ds}^{2}&=&-\left(1-\frac{2M}{r}\right)dv^{2}+2dvdr
                  +r^{2}d\Omega^{2}\\
\mbox{retarded:}\ \ {ds}^{2}&=&-\left(1-\frac{2M}{r}\right)dw^{2}-2dwdr
                  +r^{2}d\Omega^{2} \ \ .
\end{eqnarray*}

Now let us make both extensions simultaneously, then  
we will get $ds_{T-S-T}^{2}$ in the coordinates $(v,w,\theta,\phi)$:

\begin{eqnarray*}
ds^{2}_{T-S-T}=-\left(1-\frac{2M}{r}\right)^{-1}dvdw+
                (r-2M)^{2}d\Omega^{2} \ \ ,
\end{eqnarray*}

where r is determined implicitly by $\frac{1}{2}(v-w)=2Mln|r|-r$.
For constant $\theta$ and $\phi$ the corresponding two-space is 
conformally flat which can be seen by defining $x=\frac{1}{2}(v-w)$
and $t=\frac{1}{2}(v+w)$. This two-space will be invariant under
$v\rightarrow v^{\prime}=v^{\prime}(v)$ and $w\rightarrow w^{\prime}
=w^{\prime}(w)$. If we define $v^{\prime}, w^{\prime}$ as

\begin{eqnarray*}
v^{\prime}=exp\left(\frac{v+2M}{4M}\right) \ , \ \ \
w^{\prime}=exp\left(\frac{-w+2M}{4M}\right) \ \ ,
\end{eqnarray*}

and introduce $x^{\prime}=\frac{1}{2}(v^{\prime}-w^{\prime})$ and
$t^{\prime}=\frac{1}{2}(v^{\prime}+w^{\prime})$ we find the maximal
extended line element and call it dual Kruskal line element (where
we have included now the integration constant $c=2Mln|2M|$):

\begin{eqnarray*}
{ds}_{T-S-T}^{2}=\frac{32M^{3}}{2M-r}exp\left(\frac{r-2M}{2M}\right)
(-d{t^{\prime}}^{2}+d{x^{\prime}}^{2})+(r(t^{\prime},x^{\prime})-2M)^{2}
d\Omega^{2} \ \ .
\end{eqnarray*}

The Kruskal diagram is given by:

\begin{eqnarray*}
(t^{\prime})^{2}-(x^{\prime})^{2}=\frac{r}{2M}\ \ exp\left(
\frac{-r+2M}{2M}\right)
\end{eqnarray*}

At this point it would be reasonable to consider the original
Kruskal line element given by 

\begin{eqnarray*}
{ds}^{2}=\frac{32M^{3}}{r}exp\left(\frac{-r}{2M}\right)
(-d{t^{\prime}}^{2}+d{x^{\prime}}^{2})+r(t^{\prime},x^{\prime})^{2}
d\Omega^{2} \ \ .
\end{eqnarray*}

Kruskal's choice of the functions $v^{\prime},w^{\prime}$ 
was $v^{\prime}=exp(v/4M)$, $w^{\prime}=-exp(-w/4M)$. The corresponding
diagram is given by:

\begin{eqnarray*}
(t^{\prime})^{2}-(x^{\prime})^{2}=-(\frac{r}{2M}-1)
exp\left(\frac{r}{2M}\right)
\end{eqnarray*}

We observe that we can map the original solution to the dual via:

\begin{eqnarray*}
r\rightarrow r-2M, \ \ \ \ \ M\rightarrow -M \ \ .
\end{eqnarray*}

We have found a new background which solves the background field 
equations, that is singular at $r=2M$ and has a horizon at $r=0$.
In particular T-S-T interchanges two asymptotically flat space-times 
both of which have a maximal extension. T-S-T duality interchanges
region I ($r>2M$) with region V ($r<0$) while region II ($0<r<2M$) 
is transformed 
to itself (analog IV, VI are interchanged and III is transformed to itself).
Further, we can conclude from our analysis above that a test-string can not
distinguish if he moves on a ingoing radial null geodesic into the 
black hole or if he leaves a white hole in the dual geometry. That offers
the possibility that a test-string does not see the singularity. From 
M$\rightarrow-$M follows that the dual mass measured by an observer
in region I is negativ.  


\begin{figure}[h]
\setlength{\unitlength}{1cm}
\begin{center}
\begin{picture}(4,4)
\psfig{file=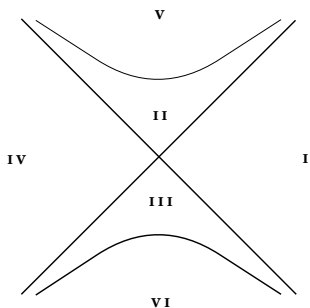 ,width = 4 cm}
\end{picture}\par
\caption{Extended space-time}
\end{center}
\end{figure}

All these facts are similar to the 2D case 
discussed in [\ref{giveon}]. In the case of the 2D black hole
one finds the positiv-mass black hole is dualized to a
negative-mass solution, i.e. there is a mapping under duality
of the asymptotically flat region to the region `beyond' 
the singularity. Both regions
are incorporated as different sectors of a single exact conformal 
field theory. Note that our discussed solutions are exact only in the 
leading order of $\alpha^{\prime}$. Finally one has to keep in mind,
although the physical equivalence of the dual solutions is expected
to hold whenever the symmetry on which the dualization is based is 
compact, their equivalence for non-compact 
symmetries is not quite clear until now. About exactness in the compact 
and non-compact case see for instance [\ref{Q3},\ref{roc},\ref{kirt}].\\

\hspace{0.5cm}

{\bf Acknowledgement:} I would like to thank D.L\"ust for  discussions.

\hspace{0.5cm}

%
%

\section*{References}
\begin{enumerate}
\item
\label{bakas}
I.Bakas, Space Time Interpretation of S-Duality and Supersymmetry
Violations of T-Duality, Phys.Lett.{\bf B343} (1995) 103, hep-th/9410104;
\item
\label{ehlers}
J.Ehlers, in ``Les Theories Relativistes de la Gravitation'', CNRS, Paris
(1959); R.Geroch, J. Math. Phys. 12 (1971) 918;
\item
\label{witt}
E.Witten, Phys.Rev.{\bf D44} (1991) 314;
\item
\label{giveon}
A.Giveon, Mod.Phys.Lett.{\bf A6} (1991) 2843-2854;\\
R.Dijkgraaf, E.Verlinde and H.Verlinde, Nucl.Phys.{\bf B371} (1992) 269;\\  
E.Kiritsis, Mod.Phys.Lett.{\bf A6} (1991) 2871;
\item
\label{luest}
D.L\"ust, preprint CERN-TH.6850/93, in Proceedings to 4th Hellenic School
on Elementary Particle Physics, Corfu (1992), hep-th/9303175;
\item
\label{Q1}
X.C.de la Ossa, F.Quevedo, Nucl.Phys.{\bf B403} (1993) 377, hep-th/9210021;
\item
\label{Q2}
F.Quevedo, Abelian and Nonabelian Dualities in String Backgrounds, in
          Erice 1992, Proceedings: From Superstrings to Supergravity 21-31,
          hep-th/9305055;
\item
\label{Q3}
C.P.Burgess, R.C.Meyrs, F.Quevedo, Nucl.Phys.{\bf B442} (1995) 97, 
hep-th/9411195;
\item
\label{Q4}
C.P.Burgess, R.C.Meyrs, F.Quevedo, hep-th/9508092;
\item
\label{giveon2}
A.Giveon, M.Porati and E.Rabinovici, Phys.Rep.244 (1994) 77, hep-th/9401139;
\item
\label{perry}
C.G.Callan, D.Friedan, E.J.Martinec und M.J.Perry, Nucl.Phys.{\bf B262}
(1985) 593;
\item
\label{hawking}
S.W.Hawking, G.F.R.Ellis: The large scale structure of space-time,
Cambridge Monographs on Mathematical Physics;
\item
\label{lu}
A.Font, L.Iba$\tilde{n}$ez, D.L\"ust and F.Quevedo, Phys.Lett.
{\bf B245} (1990) 35;
\item
\label{roc}
M.Ro$\breve{c}$ek, E.Verlinde, Nucl.Phys.{\bf B373} (1992) 630;
\item
\label{kirt}
A.Giveon, E.Kiritsis, Nucl.Phys.{\bf B411} (1994) 487;\\
E.Kiritsis, Nucl.Phys.{\bf B405} (1993) 109;
\end{enumerate}

\end{document}